\documentclass[openacc]{rsproca_new}%%%%where rsproca is the template name
\usepackage{amsmath}
\usepackage{float}
\begin{document}

%%%% Article title to be placed here
\title{Analytic and numerical solutions to the seismic wave equation in continuous media}

\author{%%%% Author details
S. J. Walters$^{1}$, L. K. Forbes and A. M. Reading}

%%%%%%%%% Insert author address here
\address{$^{1}$Mathematics and Physics, University of Tasmania}

%%%% Subject entries to be placed here %%%%
\subject{Geophysics, Applied Mathematics}

%%%% Keyword entries to be placed here %%%%

%%%% Insert corresponding author and its email address}
\corres{Stephen J Walters\\
\email{sjw1@utas.edu.au}}

%%%% Abstract text to be placed here %%%%%%%%%%%%
\begin{abstract}
This paper presents two approaches to mathematical modelling of a synthetic seismic pulse, and a comparison between them. First, a new analytical model is developed in two-dimensional Cartesian coordinates. Combined with an initial condition of sufficient symmetry, this provides a valuable check for the validity of the numerical method that follows. A particular initial condition is found which allows for a new closed-form solution. A numerical scheme is then presented which combines a spectral (Fourier) representation for displacement components and wave-speed parameters, a fourth order Runge-Kutta integration method, and an absorbing boundary layer. The resulting large system of differential equations is solved in parallel on suitable enhanced performance desktop hardware in a new software implementation. This provides an alternative approach to forward modelling of waves within isotropic media which is efficient, and tailored to rapid and flexible developments in modelling seismic structure, for example, shallow depth environmental applications. Visual comparisons of the analytic solution and the numerical scheme are presented.
\end{abstract}
%%%%%%%%%%%%%%%%%%%%%%%%%%%

%%%%%%%%%% Insert the texts which can accomdate on firstpage in the tag "fmtext" %%%%%

\begin{fmtext}
\end{fmtext}

\maketitle
%%%%%%%%%%%%%%% End of first page %%%%%%%%%%%%%%%%%%%%%
\section{Introduction}
Wave propagation in elastic media, and the accurate capture of the resulting time series of velocity variation with time for a given point, are of significant value in the subsequent determination of hidden subterranean structure.  Through these processes, seismic waves potentially yield useful information on material properties and dimensions of the structures, and other bodies, comprising the buried parts of the Earth  \cite{aki,kennett}.

Seismic wave propagation has attracted interest for many decades.  An early focus was on the propagation of surface waves, being the highest amplitude and therefore most destructive disturbances resulting from shallow earthquakes. Lamb investigated the development of an initial impulse to the surface of an elastic half space (often referred to as Lamb's problem \cite{lamb}). During the twentieth century, this was further investigated, notably by Garvin \cite{garvin}. Kausel \cite{kausel} recast the problem and solution into a relatively simple form. Analytic solutions to the seismic equations in an infinite space have also been considered, although receiving less attention than Lamb's problem. Solutions have been found for simple initial conditions such as a step-function or a Dirac delta function \cite{gosselin,carcione}. Semi-analytic solutions have also been found in terms of Green's functions \cite{diaz}.

Using a numerical approach (see \cite{igel} for descriptions of the six most common methods), a number of advanced wave propagation and waveform modelling codes exist that were developed for complex Earth models and global seismology applications \cite{fichtner,maeda,koma}. However, as seismic methods gain increasing usage in environmental and other near-surface applications (e.g. \cite{sens,tsai,winberry}), the research imperative has emerged for an efficient waveform propagation code tailored to simple, readily adaptable, structures. Such code would run at high resolution on modest hardware and hence be of wide, practical value where the application calls for an agile implementation. Questions of interest include the optimisation and placement of seismic sensors to resolve information about an underlying ice sheet and its interface with the bedrock beneath.  

In this contribution, following a summary of relevant theory, we present a new analytic solution for the time-dependent propagation of seismic signals in an infinite 2D space. Use of an infinite space allows freedom in the choice of initial conditions to suit the application of interest.  A further advantage is the relative simplicity of solutions for the infinite space which facilitates testing of numerical codes for propagation within a medium.  Choice of particular initial conditions leads to particular solutions, and we found one case which has a closed form solution.  We also present a novel numerical implementation for media with smoothly varying wave-speed parameters. This numerical scheme is validated against the analytical solution, and shows the wave propagation and synthetic seismic waveforms for simple structures within a continuous isotropic medium.  

\section{Theoretical Background}
In this section we briefly review the equations of motion for a linear, isotropic, elastic system. We begin with Cauchy's momentum equation (that is, Newton's second law) in index notation:

\begin{eqnarray}
\rho\frac{\partial^2 u_i}{\partial t^2}=\partial_j \tau_{ij}+f_i,
\label{momentum1}
\end{eqnarray}
where $\rho$ is the density of the medium, $u_i$ are the components of the displacement from equilibrium at each point, $\partial_j\equiv \partial/\partial x_j$ is the derivative operator, $\tau_{ij}$ is the stress tensor and $f_i$ are the components of any forcing terms (e.g. gravity). In this paper we ignore such forcing terms. Also, we work exclusively in Cartesian coordinates, so all coordinates may be written as lower indices. We also adopt the Einstein summation convention, whereby a repeated index in a single term implies summation over that index. Assuming a linear, isotropic stress-strain relation (Hooke's law), the stress tensor may be defined as
\begin{eqnarray}
\tau_{ij}=\lambda \delta_{ij} e_{kk}+2\mu e_{ij},
\label{tij}
\end{eqnarray}
where $\lambda$ and $\mu$ are the Lam\'e parameters, $\delta_{ij}$ is the Kronecker delta symbol, and Cauchy's strain tensor $e_{ij}$ is defined as
\begin{eqnarray}
e_{ij}=\tfrac{1}{2}(\partial_i u_j+\partial_j u_i).
\label{eij}
\end{eqnarray}
The generalized Hooke's law (\ref{tij}) is given in \cite{aki}, p.34, and we observe that the Lam\'e terms $\lambda$ and $\mu$ may be functions of position within the medium. Substituting (\ref{eij}) into (\ref{tij}), we can write the stress tensor as 
\begin{eqnarray}
\tau_{ij}=\lambda \delta_{ij} \partial_k u_k+\mu (\partial_i u_j+\partial_j u_i).
\label{tij2}
\end{eqnarray}
We may now substitute (\ref{tij2}) into (\ref{momentum1}) to write out the displacement wave equation:
\begin{eqnarray}
\rho\frac{\partial^2 u_i}{\partial t^2}&=&\partial_j[\lambda \delta_{ij} \partial_k u_k+\mu (\partial_i u_j+\partial_j u_i)].\nonumber
\end{eqnarray}
In compact notation, this may be written as
\begin{eqnarray}
\rho u_{i,tt}&=&\lambda_{,i}u_{k,k}+\mu_{,j}(u_{j,i}+u_{i,j})+(\lambda+\mu)u_{j,ij}+\mu u_{i,jj}.
\label{wave1}
\end{eqnarray}
Here we have used the comma notation for derivatives, in which indices following the comma indicate differentiation with respect to the indicated coordinate (e.g. $u_{i,jj}\equiv \partial^2 u_i/\partial x_j^2)$.

In this paper, we will consider two-dimensional systems. Designating the two  components of displacement with upper indices for readability, $u^X$ and $u^Y$, we write (\ref{wave1}) as
\begin{eqnarray}
\rho\frac{\partial^2 u^X}{\partial t^2}&=&\frac{\partial\lambda}{\partial x}\big(\frac{\partial u^X}{\partial x}+\frac{\partial u^Y}{\partial y}\big)+2\frac{\partial\mu}{\partial x}\frac{\partial u^X}{\partial x}+\frac{\partial\mu}{\partial y}\big(\frac{\partial u^Y}{\partial x}+\frac{\partial u^X}{\partial y}\big)\nonumber\\
&+&(\lambda+2\mu)\frac{\partial^2 u^X}{\partial x^2}+(\lambda+\mu)\frac{\partial^2 u^Y}{\partial x \partial y}+\mu\frac{\partial^2 u^X}{\partial y^2}\nonumber\\
\rho\frac{\partial^2 u^Y}{\partial t^2}&=&\frac{\partial\lambda}{\partial y}\big(\frac{\partial u^X}{\partial x}+\frac{\partial u^Y}{\partial y}\big)+2\frac{\partial\mu}{\partial y}\frac{\partial u^Y}{\partial y}+\frac{\partial\mu}{\partial x}\big(\frac{\partial u^Y}{\partial x}+\frac{\partial u^X}{\partial y}\big)\nonumber\\
&+&(\lambda+2\mu)\frac{\partial^2 u^Y}{\partial y^2}+(\lambda+\mu)\frac{\partial^2 u^X}{\partial x \partial y}+\mu\frac{\partial^2 u^Y}{\partial x^2}.
\label{full2d}
\end{eqnarray}
\section{Analytical Model}
\label{sec_analytic}
In order to check the accuracy of any numerical model, and in particular the method which we present in Section \ref{sec_numeric}, it is valuable to derive an exact non-trivial solution for the system.
In order to find an exact solution, we may consider a particular case of (\ref{full2d}) for the parameters and boundary conditions. As we are trying to find an exactly solvable system, we consider the case where the Lam\'e parameters $\lambda$ and $\mu$ and the density $\rho$ are constant throughout the space. Equations \ref{full2d} then simplify to
\begin{eqnarray}
\rho\frac{\partial^2 u^X}{\partial t^2}&=&(\lambda+2\mu)\frac{\partial^2 u^X}{\partial x^2}+(\lambda+\mu)\frac{\partial^2 u^Y}{\partial x \partial y}+\mu\frac{\partial^2 u^X}{\partial y^2}\nonumber\\
\rho\frac{\partial^2 u^Y}{\partial t^2}&=&(\lambda+2\mu)\frac{\partial^2 u^Y}{\partial y^2}+(\lambda+\mu)\frac{\partial^2 u^X}{\partial x \partial y}+\mu\frac{\partial^2 u^Y}{\partial x^2}.
\label{constant2d}
\end{eqnarray}
For the boundary conditions, we assume an infinite medium in which any initial disturbance never reaches the boundary at infinity, so that $u^X, u^Y \rightarrow 0$ as $x,y\rightarrow\pm\infty$. Additionally, we specify an initial disturbance to the displacement: $u^X(x,y,0)=f(x,y)$ and $u^Y(x,y,0)=g(x,y)$ for some specified initial displacement functions $f$ and $g$. The medium is set to be initially at rest: $\partial u^X/\partial t=\partial u^Y/\partial t=0$ at $t=0$.

We now seek a solution using Fourier transforms. The transform and inverse transform for the displacement are
\begin{eqnarray}
\hat U^X(\omega_1,\omega_2;t)=\frac{1}{2\pi}\int_{-\infty}^{\infty}\int_{-\infty}^{\infty}u^X(x,y,t)e^{-i\omega_1x-i\omega_2y}\text d x\text d y\nonumber\\
u^X(x,y,t)=\frac{1}{2\pi}\int_{-\infty}^{\infty}\int_{-\infty}^{\infty}\hat U^X(\omega_1,\omega_2;t)e^{i\omega_1x+i\omega_2y}\text d \omega_1\text d \omega_2,\nonumber
\end{eqnarray}
and similar forms for $u^Y$.
The known initial conditions $f$ and $g$ also have Fourier Transforms
\begin{eqnarray}
\hat F(\omega_1,\omega_2)=\frac{1}{2\pi}\int_{-\infty}^{\infty}\int_{-\infty}^{\infty}f(x,y)e^{-i\omega_1x-i\omega_2y}\text d x\text d y\nonumber\\
f(x,y)=\frac{1}{2\pi}\int_{-\infty}^{\infty}\int_{-\infty}^{\infty}\hat F(\omega_1,\omega_2)e^{i\omega_1x+i\omega_2y}\text d \omega_1\text d \omega_2,\nonumber
\end{eqnarray}
and similar forms for $g$.

Application of these Fourier Transforms to the partial differential equations (\ref{constant2d}) and initial conditions leads to a system of two ordinary differential equations in the Fourier space:

\begin{eqnarray}
\rho (\hat U^X)^{''}=-(\omega_1^2(\lambda+2\mu)+\omega_2^2 \mu)\hat U^X-(\lambda+\mu)\omega_1 \omega_2\hat U^Y\nonumber\\
\rho (\hat U^Y)^{''}=-(\omega_2^2(\lambda+2\mu)+\omega_1^2 \mu)\hat U^Y-(\lambda+\mu)\omega_1 \omega_2\hat U^X\nonumber\\
\hat U^X(0)=\hat F, \hat U^Y(0)=\hat G,\frac{d\hat U^X}{dt}(0)=0, \frac{d\hat U^Y}{dt}(0)=0, \nonumber
\end{eqnarray}
where the derivatives denote differentiation with respect to time $t$.

Following some algebra in which we rearrange to eliminate the transformed variable $\hat U^Y$ in terms of $\hat U^X$, we obtain the fourth order ordinary differential equation for $\hat U^X$ (\ref{ux4th}). This equation is constant coefficient and linear:
\begin{eqnarray}
\rho^2(\hat U^X)^{IV}+\rho(A+C)(\hat U^X)''+(AC-B^2)\hat U^X=0,
\label{ux4th}
\end{eqnarray}
where
\begin{eqnarray}
A&=&\omega_1^2(\lambda+2\mu)+\omega_2^2\mu\nonumber\\
B&=&(\lambda+\mu)\omega_1\omega_2\nonumber\\
C&=&\omega_2^2(\lambda+2\mu)+\omega_1^2\mu.\nonumber
\end{eqnarray}
It is now straightforward to show that
\begin{eqnarray}
\hat U^X(t)&=&P_1 \text{exp}\bigg[it\sqrt{(\omega_1^2+\omega_2^2)\mu/\rho}\bigg]+P_2 \text{exp}\bigg[-it\sqrt{(\omega_1^2+\omega_2^2)\mu/\rho}\bigg]\nonumber\\
&+&Q_1 \text{exp}\bigg[it\sqrt{(\omega_1^2+\omega_2^2)(\lambda+2\mu)/\rho}\bigg]+Q_2 \text{exp}\bigg[-it\sqrt{(\omega_1^2+\omega_2^2)(\lambda+2\mu)/\rho}\bigg]\nonumber\\
\hat U^Y(t)&=&-P_1\frac{\omega_1}{\omega_2} \text{exp}\bigg[it\sqrt{(\omega_1^2+\omega_2^2)\mu/\rho}\bigg]-P_2\frac{\omega_1}{\omega_2} \text{exp}\bigg[-it\sqrt{(\omega_1^2+\omega_2^2)\mu/\rho}\bigg]\nonumber\\
&+&Q_1\frac{\omega_2}{\omega_1} \text{exp}\bigg[it\sqrt{(\omega_1^2+\omega_2^2)(\lambda+2\mu)/\rho}\bigg]+Q_2\frac{\omega_2}{\omega_1} \text{exp}\bigg[-it\sqrt{(\omega_1^2+\omega_2^2)(\lambda+2\mu)/\rho}\bigg]\nonumber,
\end{eqnarray}
where the constants $P_1,P_2,Q_1$ and $Q_2$ are found from the Fourier transform of the initial conditions, $\hat F$ and $\hat G$. Solving for these initial conditions gives the complete solution
\begin{eqnarray}
\hat U^X(t)&=&\frac{\omega_2(\hat F \omega_2-\hat G \omega_1)}{\omega_1^2+\omega_2^2}\cos \bigg(t\sqrt{\frac{\mu}{\rho}(\omega_1^2+\omega_2^2)}\bigg)+\frac{\omega_1(\hat F \omega_1+\hat G \omega_2)}{\omega_1^2+\omega_2^2}\cos \bigg(t\sqrt{\frac{\lambda+2\mu}{\rho}(\omega_1^2+\omega_2^2)}\bigg)\nonumber\\
\hat U^Y(t)&=&-\frac{\omega_1(\hat F \omega_2-\hat G \omega_1)}{\omega_1^2+\omega_2^2}\cos \bigg(t\sqrt{\frac{\mu}{\rho}(\omega_1^2+\omega_2^2)}\bigg)+\frac{\omega_2(\hat F \omega_1+\hat G \omega_2)}{\omega_1^2+\omega_2^2}\cos \bigg(t\sqrt{\frac{\lambda+2\mu}{\rho}(\omega_1^2+\omega_2^2)}\bigg)\nonumber\\
\label{completesol}
\end{eqnarray}
in the Fourier-transformed space.
\subsubsection{Initial Conditions}
It remains to determine the transformed quantities $\hat F$ and $\hat G$ in (\ref{completesol}), through an appropriate  choice of initial conditions $f(x,y)$ and $g(x,y)$ in the physical space. Here we choose
\begin{eqnarray}
u^X(x,y,0)\equiv f(x,y)&=&F_0\frac{\partial}{\partial x}(e^{-(x^2+y^2)/a^2})\nonumber\\
&=&-F_0\frac{2x}{a^2}e^{-(x^2+y^2)/a^2}\nonumber\\
u^Y(x,y,0)\equiv g(x,y)&=&G_0\frac{\partial}{\partial y}(e^{-(x^2+y^2)/a^2})\nonumber\\
&=&-G_0\frac{2y}{a^2}e^{-(x^2+y^2)/a^2},
\label{ic1}
\end{eqnarray}
for given constants $F_0, G_0, a$.
This implies that the initial conditions in the Fourier space are
\begin{eqnarray}
\hat F(\omega_1,\omega_2)&=&-\frac{2F_0}{a^2}\frac{1}{2\pi}\int_{-\infty}^{\infty}\int_{-\infty}^{\infty}x e^{-(x^2+y^2)/a^2-i\omega_1 x-i\omega_2 y}\text d x \text d y\nonumber\\
&=&\frac{i F_0 a^2 \omega_1}{2} e^{\frac{1}{4} a^2 \left(\omega_1^2+\omega_2^2\right)}\nonumber\\
\hat G(\omega_1,\omega_2)&=&-\frac{2G_0}{a^2}\frac{1}{2\pi}\int_{-\infty}^{\infty}\int_{-\infty}^{\infty}y e^{-(x^2+y^2)/a^2-i\omega_1 x-i\omega_2 y}\text d x \text d y\nonumber\\
&=&\frac{i G_0 a^2 \omega_2}{2} e^{\frac{1}{4} a^2 \left(\omega_1^2+\omega_2^2\right)}\nonumber
\label{ic2}
\end{eqnarray}
Inserting these initial conditions into (\ref{completesol}) gives the following equations in Fourier space:
\begin{eqnarray}
\hat U^X(t)&=&\frac{i a^2 \omega_1 \omega_2^2}{2(\omega_1^2+\omega_2^2)} e^{\frac{1}{4} a^2 (\omega_1^2+\omega_2^2)}(F_0-G_0)\cos \bigg(t\sqrt{\frac{\mu}{\rho}(\omega_1^2+\omega_2^2)}\bigg))\nonumber\\
&+&\frac{i a^2 \omega_1}{2}\frac{F_0 \omega_1^2+G_0  \omega_2^2}{\omega_1^2+\omega_2^2}e^{\frac{1}{4} a^2 \left(\omega_1^2+\omega_2^2\right)}\cos \bigg(t\sqrt{\frac{\lambda+2\mu}{\rho}(\omega_1^2+\omega_2^2)}\bigg)\nonumber\\
\hat U^Y(t)&=&\frac{i a^2 \omega_1^2 \omega_2}{2(\omega_1^2+\omega_2^2)} e^{\frac{1}{4} a^2 (\omega_1^2+\omega_2^2)}(G_0-F_0)\cos \bigg(t\sqrt{\frac{\mu}{\rho}(\omega_1^2+\omega_2^2)}\bigg))\nonumber\\
&+&\frac{i a^2 \omega_2}{2}\frac{F_0 \omega_1^2+G_0  \omega_2^2}{\omega_1^2+\omega_2^2}e^{\frac{1}{4} a^2 \left(\omega_1^2+\omega_2^2\right)}\cos \bigg(t\sqrt{\frac{\lambda+2\mu}{\rho}(\omega_1^2+\omega_2^2)}\bigg)\nonumber\\
\label{specialsol}
\end{eqnarray}

After some algebra, the solution in physical space is found from the inverse Fourier transformation of (\ref{specialsol}) to be
\begin{eqnarray}
u^X(x,y,t)&=&-\tfrac{1}{2}a^2\frac{x}{r^3}(x^2 F_0+y^2 G_0)B_2(r,t)-\tfrac{1}{2}a^2(F_0-G_0)\bigg\{\frac{x y^2}{r^3}A_2(r,t)\nonumber\\
&+&\frac{2x(x^2-3y^2)}{r^5}\big[A_1(r,t)-B_1(r,t)\big]-\frac{x(x^2-3y^2)}{r^4}\big[A_3(r,t)-B_3(r,t)\big]\bigg\}\nonumber\\
u^Y(x,y,t)&=&-\tfrac{1}{2}a^2\frac{y}{r^3}(x^2 F_0+y^2 G_0)B_2(r,t)-\tfrac{1}{2}a^2(F_0-G_0)\bigg\{\frac{x^2 y}{r^3}A_2(r,t)\nonumber\\
&+&\frac{2y(y^2-3x^2)}{r^5}\big[A_1(r,t)-B_1(r,t)\big]-\frac{y(y^2-3x^2)}{r^4}\big[A_3(r,t)-B_3(r,t)\big]\bigg\},
\label{exact1}
\end{eqnarray}
where we have introduced the usual radial distance, $r=\sqrt{x^2+y^2}$. For readability, we have also introduced the following functions of $r,t$:
\begin{eqnarray}
A_1(r,t)&=&\int_0^\infty e^{-\tfrac{1}{4}a^2 k^2} \cos(k t \sqrt{\alpha})J_1(k r)\text d k\nonumber\\
A_2(r,t)&=&\int_0^\infty e^{-\tfrac{1}{4}a^2 k^2} \cos(k t \sqrt{\alpha})k^2 J_1(k r)\text d k\nonumber\\
A_3(r,t)&=&\int_0^\infty e^{-\tfrac{1}{4}a^2 k^2} \cos(k t \sqrt{\alpha})k J_0(k r)\text d k\nonumber\\
B_1(r,t)&=&\int_0^\infty e^{-\tfrac{1}{4}a^2 k^2} \cos(k t \sqrt{\beta})J_1(k r)\text d k\nonumber\\
B_2(r,t)&=&\int_0^\infty e^{-\tfrac{1}{4}a^2 k^2} \cos(k t \sqrt{\beta})k^2 J_1(k r)\text d k\nonumber\\
B_3(r,t)&=&\int_0^\infty e^{-\tfrac{1}{4}a^2 k^2} \cos(k t \sqrt{\beta})k J_0(k r)\text d k
\label{integrals}
\end{eqnarray}
and $\alpha=\mu/\rho$ and $\beta=(\lambda+2\mu)/\rho$ are the squares of the azimuthal and radial wave speeds respectively. The functions $J_0$ and $J_1$ are the zeroth-order and first-order Bessel functions of the first kind. We now have an exact solution which can be plotted over time, and will be used to inspect the accuracy of the numerical approach in section \ref{sec_numeric}. This analytical solution requires the numerical evaluation of integrals in the $A$ and $B$ functions. These can be evaluated extremely rapidly and with any desired accuracy using numerical integration, such as the trapezoidal method or Gauss-Legendre quadrature.
\subsubsection{Closed-Form Solution}
Following some experimentation with various initial conditions, a particular initial displacement was found to yield a closed-form solution. In addition to being useful in testing numerical schemes for accuracy, a closed-form solution is of theoretical and mathematical interest. These elasto-dynamical equations are of sufficient complexity that non-trivial closed-form solutions may be elusive.
In the previous section, the chosen initial condition was the derivative of a smoothed Gaussian. This resulted in an analytic solution in terms of the integrals given in (\ref{integrals}). Here, we instead choose the following initial condition for the displacement:
\begin{eqnarray}
u^X(x,y,0)\equiv f(x,y)&=&\frac{\partial}{\partial x}(\frac{1}{(x^2+1)(y^2+1)})\nonumber\\
&=&-\frac{2x}{(x^2+1)^2(y^2+1)}\nonumber\\
u^Y(x,y,0)\equiv g(x,y)&=&\frac{\partial}{\partial y}(\frac{1}{(x^2+1)(y^2+1)})\nonumber\\
&=&-\frac{2y}{(x^2+1)(y^2+1)^2}.
\end{eqnarray}
Following the Fourier transform of this initial condition, and solution of equations (\ref{completesol}), the inverse Fourier transform yields the solutions
\begin{eqnarray}
u^X(x,y,t)&=&\frac{1}{4 i} \int_0^{2\pi} \cos \phi\big[\big(\frac{1}{A^+}\big)^3+\big(\frac{1}{A^-}\big)^3\big]\text d\phi\nonumber\\
u^Y(x,y,t)&=&\frac{1}{4 i} \int_0^{2\pi} \sin \phi\big[\big(\frac{1}{A^+}\big)^3+\big(\frac{1}{A^-}\big)^3\big]\text d\phi, \nonumber\\
\text{where  } A^{\pm}&=&\lvert \cos\phi\rvert+\lvert \sin\phi\rvert+i x\cos\phi+i y\sin\phi \pm \beta t.
\label{cf1}
\end{eqnarray}
Again, $\beta=(\lambda+2\mu)/\rho$ is the square of the radial wave speed. Consideration of the four quadrants is used to eliminate the absolute value signs, yielding the sum of eight integrals:
\begin{eqnarray}
u^X(x,y,t)&=&\frac{1}{4 i} \sum_{m=\pm 1} \sum_{n=\pm 1} \sum_{p=\pm 1} \int_0^{\pi/2}\frac{m  \cos \phi}{( \cos \phi+ \sin \phi+i (m x  \cos \phi+p y  \sin \phi+n \beta t))^3}\text d\phi\nonumber\\
u^Y(x,y,t)&=&\frac{1}{4 i} \sum_{m=\pm 1} \sum_{n=\pm 1} \sum_{p=\pm 1} \int_0^{\pi/2}\frac{m  \cos \phi}{( \cos \phi+ \sin \phi+i (m y  \cos \phi-p x  \sin \phi+n \beta t))^3}\text d\phi. \nonumber
\end{eqnarray}
It may be seen that these terms occur as four pairs of complex conjugates, so that the resulting displacement is entirely real. However, expanding in this way results in a cumbersome number of terms, so the integrals are performed as they are. The solutions are included in an appendix. These solutions are easily coded in Mathematica, and the results are shown alongside the corresponding numerical solutions in Fig.\ref{fig5} below.

\section{Numerical Approach}
\label{sec_numeric}
The modelling of the seismic equations (\ref{full2d}) for a more general initial condition, or for more complicated functions for density and Lam\'e parameters requires an efficient and accurate numerical method. The advantages and disadvantages of the most common methods are described in \cite{igel}. We present here an alternative method, developed in the context of fluid flow studies \cite{walters}. This is a spectral method, whereby variables are represented as weighted sums of analytic functions. As with the pseudospectral method \cite{igel} the spatial derivatives are generated analytically from the spectral basis functions. This provides two advantages over a finite difference scheme in that derivatives are known and cached beforehand, which is an otherwise time consuming numerical procedure, and calculation of derivatives in this way exhibits exponential convergence \cite{boyd}, pp. 45-46. However, unlike traditional pseudospectral schemes, our method is easily parallelisable, lending itself to efficient solution over large numbers of simple processing units, as found on modern graphics cards. Thus, we get the accuracy advantages of the spectral method, combined with computational efficiency approaching that of a finite difference scheme. The main disadvantage of a spectral method such as this, is that it may perform poorly when dealing with discontinuities or very rapid changes in wave-speed parameters, unless those discontinuities are explicitly allowed for in the choice of basis functions.

The numerical calculations are limited to a rectangular region, $x_1<x<x_2$, $y_1<y<y_2$. Within this region, we choose to represent the displacement as
\begin{eqnarray}
u^X(x,y,t)&=&\sum_{m=0}^M \sum_{n=0}^N A_{mn}(t) \cos\big[\alpha_m(x-x_1)\big] \cos\big[\alpha_n(y-y_1)\big]\nonumber\\
u^Y(x,y,t)&=&\sum_{m=0}^M \sum_{n=0}^N B_{mn}(t) \cos\big[\alpha_m(x-x_1)\big] \cos\big[\alpha_n(y-y_1)\big],\label{reps}
\end{eqnarray}
where $\alpha_m=m\pi/(x_2-x_1)$ and $\alpha_n=n\pi/(y_2-y_1)$.
This particular choice of basis functions allows the displacement to have any value at any point in the region and does not force any bi-lateral symmetry. However, the orthogonality condition of the Fourier series is preserved, allowing $A_{mn}$ and $B_{mn}$ to be obtained accurately and efficiently from $u^X$ and $u^Y$. In order to solve (\ref{full2d}) numerically, the two second order equations are replaced with four first order equations, resulting in two more sets of coefficients $C_{mn}, D_{mn}$ for the velocities $\partial u^X/\partial t, \partial u^Y/\partial t$. The total system of first order equations is then

\begin{gather}
\frac{\text d }{\text d t}
 \begin{bmatrix} A_{mn} \\ B_{mn} \\ C_{mn} \\ D_{mn} \end{bmatrix}
 =
  \begin{bmatrix}
   C_{mn} \\
   D_{mn} \\
   \frac{\gamma_{mn}}{v}\int_{x1}^{x2}\int_{y1}^{y2}\ddot{u}^X \cos\big[\alpha_m(x-x_1)\big] \cos\big[\alpha_n(y-y_1)\big]\text dy\text dx \\
   \frac{\gamma_{mn}}{v}\int_{x1}^{x2}\int_{y1}^{y2}\ddot{u}^Y \cos\big[\alpha_m(x-x_1)\big] \cos\big[\alpha_n(y-y_1)\big] \text dy\text dx
   \end{bmatrix}
   \label{matrixode}
\end{gather}
To be clear, in these equations, we have taken the two second-order equations (\ref{full2d}), represented them as Fourier series, and are solving for the coefficents ($A_{mn}(t)$ etc.) via these four first order equations (\ref{matrixode}). This formulation allows for solution by explicit time-integration methods, such as the Runge-Kutta scheme used in this current work.  The symbol $\gamma_{mn}$ is due to the orthogonality condition of the Fourier series, and is $1$ if $m$ and $n$ are both zero, $2$ if exactly one of $m$ or $n$ is zero, $4$ otherwise. The volume of the region (or area in this 2D case) is $v=(x_2-x_1)(y_2-y_1)$. The values of $\ddot{u}^X$ and $\ddot{u}^Y$ at each time step are determined by the wave equation (\ref{full2d}). The derivatives in equation (\ref{full2d}) are calculated analytically from the Fourier representations (\ref{reps}). The trigonometric basis functions in (\ref{reps}) are calculated only once and cached in memory for use throughout the running of the algorithm.

It is necessary in numerical calculations to avoid non-physical reflections, which may result from the boundaries of the calculation region. To achieve this, a simple robust and efficient absorbing boundary was developed. The region of interest is surrounded by a region in which an increasing amount of damping is applied to the displacement, in proportion to the velocity of the disturbance at that point. In order to effect this, we initially added a damping term to the displacement equations (\ref{full2d}). A damping field was implemented which increased exponentially throughout the absorbing region. This was applied to the displacement at each time step according to:
\begin{eqnarray}
u^X_d(x,y)&=&u^X(x,y)-\dot u^X(x,y) D(x,y)\nonumber\\
u^Y_d(x,y)&=&u^Y(x,y)-\dot u^Y(x,y) D(x,y),\nonumber
\end{eqnarray}
where the subscript $d$ refers to the (new) damped displacement, $u^X,u^Y$ are the displacement components, calculated from the forward integration of (\ref{full2d}), $\dot u^X$ and $\dot u^Y$ are the velocity components, and $D$ is the damping field. Specifically, $D$ is zero in the region of interest, but has value $D=p_1 (e^{p_2 d}-1)$ in the boundary layer, where $d$ is the depth into the absorbing region. The optimal values of the damping parameters $p_1, p_2$ were determined using the Nelder-Mead downhill simplex method \cite{nm}. By damping in proportion to the velocity normal to the boundaries (i.e. $\dot u^X$ and $\dot u^Y$), we are effectively implementing a partial matching condition for waves of varying speeds. This will be useful in models where the wavespeed varies throughout the region.
As an alternative method, we have also implemented the perfecly matched layer (PML) approach of \cite{assi}. In this method the second-order kinematic equations (\ref{full2d}) are modified to

\begin{eqnarray}
\rho \big [ \frac{\partial^2 u^X}{\partial t^2}+(\beta^x+\beta^y)\frac{\partial u^X}{\partial t}+\beta^x\beta^y u^X \big]&=&\frac{\partial}{\partial x}\big[(\lambda+2\mu)\frac{\partial u^X}{\partial x}+\lambda \frac{\partial u^Y}{\partial y}+(\beta^y-\beta^x)w_{11}\big]\nonumber\\
&+&\frac{\partial}{\partial y}\big[\mu (\frac{\partial u^Y}{\partial x}+\frac{\partial u^X}{\partial y})+(\beta^x-\beta^y)w_{12}\big]\nonumber\\
\rho \big [ \frac{\partial^2 u^Y}{\partial t^2}+(\beta^x+\beta^y)\frac{\partial u^Y}{\partial t}+\beta^x\beta^y u^Y \big]&=&\frac{\partial}{\partial y}\big[(\lambda+2\mu)\frac{\partial u^Y}{\partial y}+\lambda \frac{\partial u^X}{\partial x}+(\beta^x-\beta^y)w_{22}\big]\nonumber\\
&+&\frac{\partial}{\partial x}\big[\mu (\frac{\partial u^X}{\partial y}+\frac{\partial u^Y}{\partial x})+(\beta^y-\beta^x)w_{21}\big].\nonumber
\label{fullpml}
\end{eqnarray}

These equations introduce six new field variables: $\beta^x(x)$ and $\beta^y(y)$ are one-dimensional functions, which are zero within the physical region, and increase throughout the damping region as
\begin{eqnarray}
\beta^x(x)=q_1(\frac{x_d-x}{x_d-x_1})^{q_2}
\end{eqnarray}
and similar for $\beta^y(y)$. The factor $q_1$ and the index $q_2$ are free parameters to be chosen. The $w_{11},w_{12},w_{21},w_{22}$ are functions of $x,y,t$ which are initially set to zero, and evolved through time according to the following equations:
\begin{eqnarray}
\frac{\partial w_{11}}{\partial t}+\beta^x w_{11}&=&(\lambda+2\mu)\frac{\partial u^X}{\partial x}\nonumber\\
\frac{\partial w_{12}}{\partial t}+\beta^y w_{12}&=&\mu\frac{\partial u^X}{\partial y}\nonumber\\
\frac{\partial w_{21}}{\partial t}+\beta^x w_{21}&=&\mu\frac{\partial u^Y}{\partial x}\nonumber\\
\frac{\partial w_{22}}{\partial t}+\beta^y w_{22}&=&(\lambda+2\mu)\frac{\partial u^Y}{\partial y}.\nonumber
\end{eqnarray}

An additional "scaling coefficient" in each direction is employed in \cite{assi}, but we found that its use led to instability and increased reflections. This problem was indicated in \cite{assi2}, along with the recommendation, which we have followed here, not to use the scaling coefficient, unless the particular problem necessitated this due to certain instabilities. We thus have another two-parameter damping layer, the parameters in this case being $q_1$ and $q_2$. Again, the downhill simplex method was used to determine the optimum values for the two parameters, for various thicknesses of the damping boundary. Table \ref{table1} shows the greatest difference between the numerical and analytic solutions at time $t=5$, for the two methods, along with the optimal values found ($p_1$ and $p_2$ for the simple method, $q_1$ and $q_2$ for the PML). This is for the $560\times560$ simulation shown in Fig. \ref{fig1} in section \ref{comparison}. Some of these 560 grid points are used for the absorbing boundary layer, from 14 to 56, as shown.

\begin{table}
\begin{center}
\caption{Effectiveness of the two types of damping layer as in Fig. \ref{fig1}. The total grid size is $560\times560$, with the damping boundary ranging from 14 to 56 grid points. The greatest absolute difference between the analytic and numerical solutions is shown for time $t=5$, when the main body of the original impulsive wave has left the region.}
\label{table1}
\begin{tabular}{l|c|c|c}
number of points&14&28&56\\
\hline
$\text{Simple Damping}$ &&&\\
$p_1$&1.924&5.4&6.088\\
$p_2$&5.528&3.42&3.52\\
\hline
difference&0.217&0.144&0.0161\\
\hline
$\text{PML Damping}$ &&&\\
$q_1$&74.0&85&44.7\\
$q_2$&1.54&2.5&1.69\\
\hline
difference&$0.0188$&0.0037&0.0004

\end{tabular}
\end{center}
\end{table}

It is clear from the results in Table. \ref{table1} that the PML boundary is far more effective at damping reflections, producing a decrease in reflections by a factor of between 10 and 40 over the simpler method. The additional cost for forward integration of the auxiliary variables resulted in no more than twice the calculation time. Because of this, we have used the PML formulation for the numerical calculations in this paper. However, we have included the simpler method as an alternative for situations where simplicity of implementation is valued over the additional effectiveness of the PML method.

In all numerical simulations, the time step-size is reduced until any changes in the solution are of an acceptable level (see \cite{atkinson} p.373-374 regarding stability of Runge-Kutta methods). In the figures presented in this paper the model is no longer visibly changing with further reduction in step-size. Additionally, by comparing with analytical solutions, the step size was reduced until differences between analytical and numerical solutions are of the same order as the errors due to the spurious boundary reflections. If further refinement is required, it is not difficult to implement a $4^{th}-5^{th}$ Runge-Kutta method which adjusts the step-size automatically, such that the fourth- and fifth-order solutions differ by less than a specified tolerance.

\section{Analytic-Numerical comparison}
\label{comparison}
\begin{figure}[ht]
\hspace*{0cm}\includegraphics[width=\textwidth]{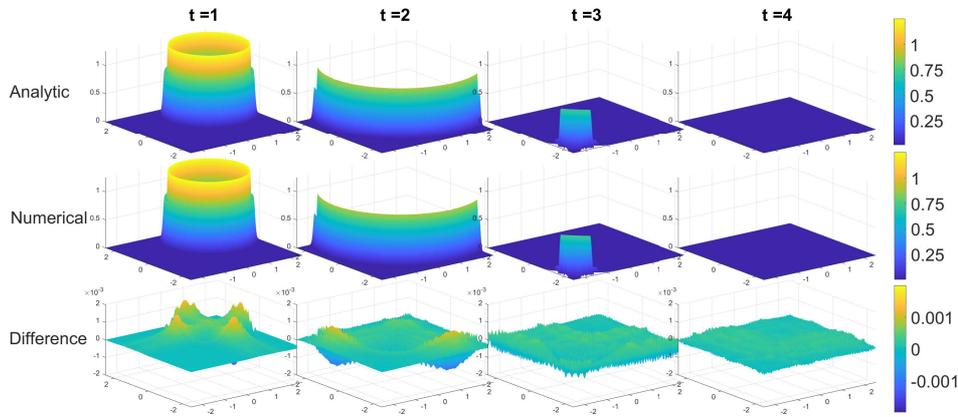}
\caption{A comparison of the expanding compressional wave at times (in seconds) from $t=1$ (left) to $t=4$ (right). The top row shows the analytic solution (\ref{exact1}), the middle row shows the output from the numerical approach of section \ref{sec_numeric} and the bottom row shows the difference between the two. The $x$ and $y$ axes are in kilometres and the $z$-axis shows the magnitude of the displacement based on a pulse with initial amplitudes $F_0=G_0=1$. In the numerical model, absorbing boundaries have been used to damp reflections.}
  %\end{center}
  \label{fig1}
\end{figure}

The numerical solution can now be compared against the analytical model as a measure of the accuracy of the numerical approach. The first check will be with an initial condition which is circularly symmetric. Using the initial condition in (\ref{ic1}), we set $F_0=G_0=1$, and $a=0.1$. The analytic solution is computed at time $t=1, t=2, t=3, t=4$ (seconds), in an $x-y$ grid of $560\times560$ regularly spaced points. A trapezoidal method was used to evaluate the integrals in (\ref{integrals}) over $k$ from $0$ to $90$ using $201$ integration points. These values for the truncation point of the integral and the number of points were chosen by increasing them until there was no change in the evaluation at double precision. The wave speeds
\begin{eqnarray}
V_p=\sqrt{\beta}=\sqrt{\frac{\lambda+2\mu}{\rho}}; V_s=\sqrt{\alpha}=\sqrt{\frac{\mu}{\rho}}
\label{wavespeeds}
\end{eqnarray}
were set at $V_p=1.5 km/s$,  $V_s=0.5 km/s$ and density $\rho=1000 kg/m^3$ throughout the space. The numerical solution was then calculated using the same parameters, and the same spatial grid. The outer $56$ points on each edge of the space were used for the PML damping layer, implemented as described in section \ref{sec_numeric}. The classic fourth-order Runge-Kutta method was used with a timestep of 0.005 seconds. Fig. \ref{fig1} shows a plot of the two methods, with the bottom row showing the difference between the two. Only the undamped region is plotted, being the inner $448\times448$ points. It may be seen from the colour scales that the difference between the two methods is limited to approximately 0.1\% of the initial amplitude.

\begin{figure}[h]
\hspace*{0cm}\includegraphics[width=\textwidth]{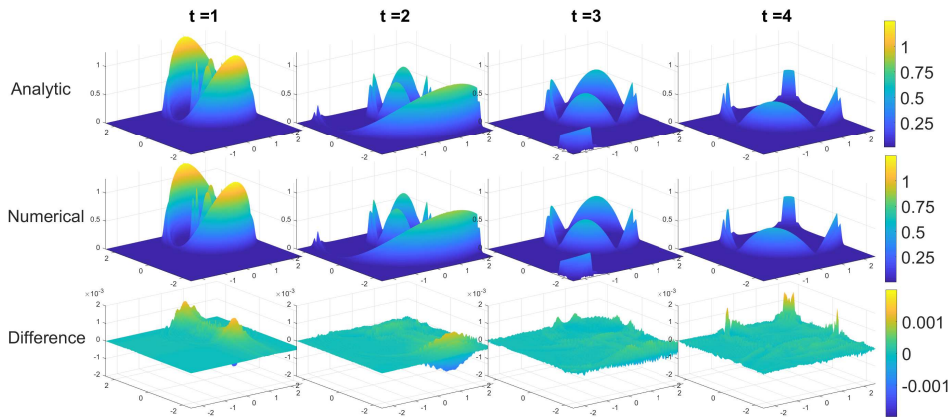}
\caption{A similar comparison of the numerical and analytical solutions as in Fig. \ref{fig1}. In this case the chosen initial condition gives rise to both fast compression waves and slower moving shear waves. In the bottom row the damped reflections from the numerical model are small but visible differences between the numerical and analytic solutions.}
  %\end{center}
  \label{fig2}
\end{figure}

A second check of the numerical scheme is shown in Fig. \ref{fig2} using the analytic solution with a different initial condition. In this case, we set initial condition parameters to be $F_0=1, G_0=0$. All other parameters remained the same as in Fig. \ref{fig1}. In this case the initial condition gives rise to both a compression wave and a slower moving shear wave. Again, the numerical solution and analytic solution were run with identical initial condition, and the difference plotted. The numerical solution shows excellent agreement with the analytic model. It can be seen from the first frame of Fig. \ref{fig2} that the amplitude of the pulse when first contacting the boundary has an amplitude in excess of $1$, while reflected amplitudes are around the $0.1\%$ level. As with Fig.\ref{fig1}, the differences may be further reduced by increasing the number of grid points, spectral modes, integration steps and the absorbing boundary thickness.

\begin{figure}[h]
\hspace*{0cm}\includegraphics[width=\textwidth]{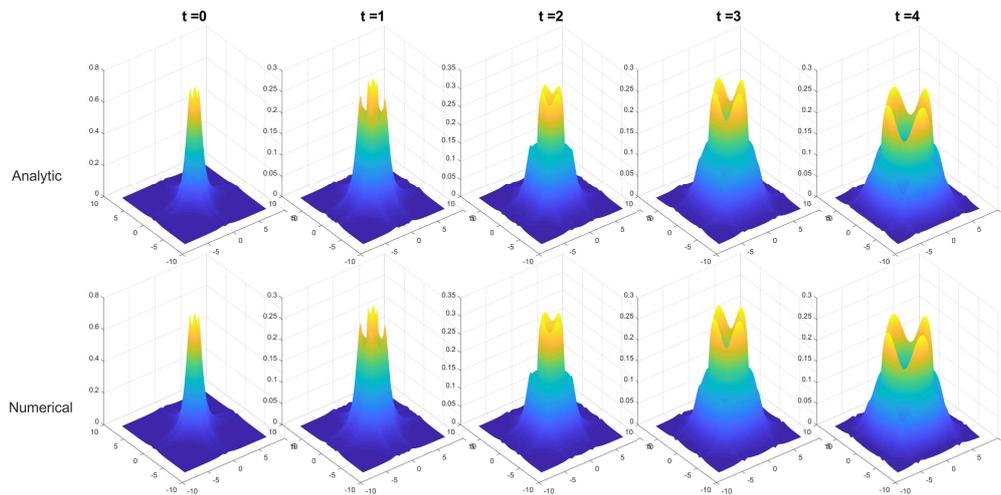}
\caption{A comparison of the numerical solution with the closed-form solution described in Section  \ref{sec_analytic}. This initial condition gives rise to compression waves only. It may be seen qualitatively that the numerical solution matches the closed-form solution well.}
  %\end{center}
  \label{fig5}
\end{figure}

A final comparison is shown in Fig \ref{fig5}. This is the closed form solution, for initial displacement
\begin{eqnarray}
u^X(x,y,0)&=&-\frac{2x}{(x^2+1)^2(y^2+1)}\nonumber\\
u^Y(x,y,0)&=&-\frac{2y}{(x^2+1)(y^2+1)^2}.\nonumber
\end{eqnarray}

The closed-form solution was calculated for the specified times, from 0 to 4, in Mathematica. The numerical solution was computed using the numerical scheme described above, marched forward in time in Fortran using parallel CUDA. The data for both was then plotted in \textsc{Matlab}. The grid size is from -8 to 8 in both directions.

\section{Variable Wavespeed Parameters}
\begin{figure}[H]
\hspace*{-0cm}\includegraphics[width=\textwidth]{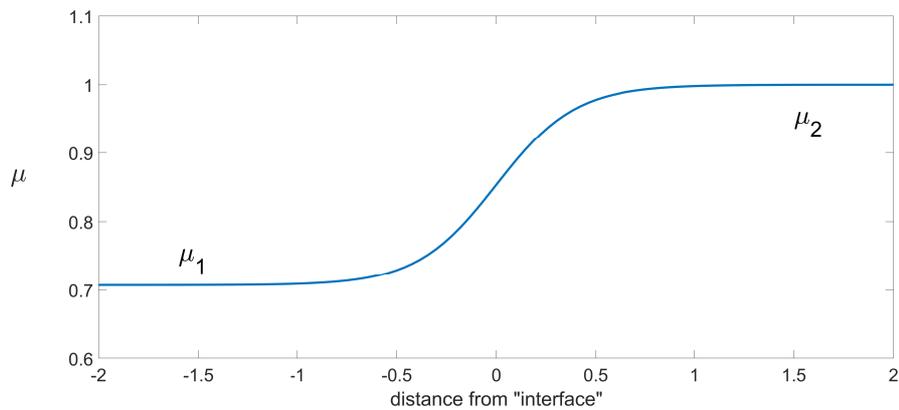}
\caption{The scaled Lam\'e parameters change smoothly between the two regions in Fig.\ref{fig3}. The change in $\mu$ is shown here according to $\mu=\mu_1+(\mu_2-\mu_1)/(1+\exp[-5 D])$, where $D$ is the distance from the line at $y=-x$. This smooth change results in a model containing refraction but not reflections (see Fig.\ref{fig3}).}
  %\end{center}
  \label{figmu}
\end{figure}

Having confidence that the numerical scheme is accurate in the case of constant Lam\'e parameters, we now move into the realm of varying parameters, for which we do not have any analytic solutions. While we cannot compare directly in these cases, we are using exactly the same numerical formulation as we used when comparing against the known constant parameter solutions shown in Figs. \ref{fig1} and \ref{fig2}. We thus have at least some confidence that such a numerical scheme is a reasonable approach to inhomogeneous media. We first show a simulation of an initially circular wavefront propagating through a medium with two different sets of wavespeeds. These wavespeeds change smoothly but rapidly across the boundary, as given below. The upper right half of the space has wavespeeds $v_s=\sqrt{\mu_1/\rho_1}=.5 km/s, v_p=\sqrt{(\lambda_1+2\mu_1)/\rho_1}=1.5 km/s$ and the lower left half has $v_s=\sqrt{\mu_2/\rho_2}=1 km/s, v_p=\sqrt{(\lambda_2+2\mu_2)/\rho_2}=2.5 km/s$. The equations really only have two independent parameters, so $\rho_1$ and $\rho_2$ were scaled out (set to 1) in the code. These scaled Lam\'e parameters were defined throughout the region as $\mu=\mu_1+(\mu_2-\mu_1)/(1+\exp[-5(x+y)])$, and similarly for $\lambda$. This change is shown in Fig.\ref{figmu} where the value of $\mu$ changes from $\mu_1$ to $\mu_2$ over the interfacial region.

\begin{figure}[H]
\hspace*{-0cm}\includegraphics[width=\textwidth]{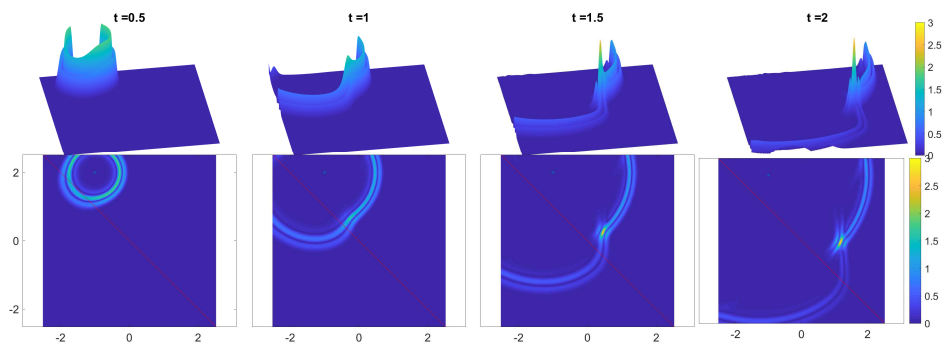}
\caption{A simulation of the propagation of an initially circular wave through a medium having two distinct sets of wave parameters, which change rapidly but smoothly across a diagonal line. The wave can be seen travelling more slowly in the upper right half of the plane. Times are given in seconds, distances are given in kilometres. The upper row shows the absolute magnitude of the displacement. The lower row shows the corresponding contour lines, along with a diagonal line showing the location of the "interface" and the star shows the centre of the initial disturbance.}
  %\end{center}
  \label{fig3}
\end{figure}

The results of this simulation can be seen in Fig. \ref{fig3}, where the wavefront is seen to be travelling faster after crossing the interface into the lower left region. A caustic has formed near the diagonal line due to the rapidly changing refractive index in the interfacial region. As with the simulations in Fig. \ref{fig1} and \ref{fig2}, a regular $x-y$ grid of $560 \times 560$ points was used. We note that a steeper parameter gradient requires a finer grid in order to avoid the Gibbs-type phenomena often associated with spectral methods.

\begin{figure}[H]
\hspace*{-2cm}\includegraphics[width=1.2\textwidth]{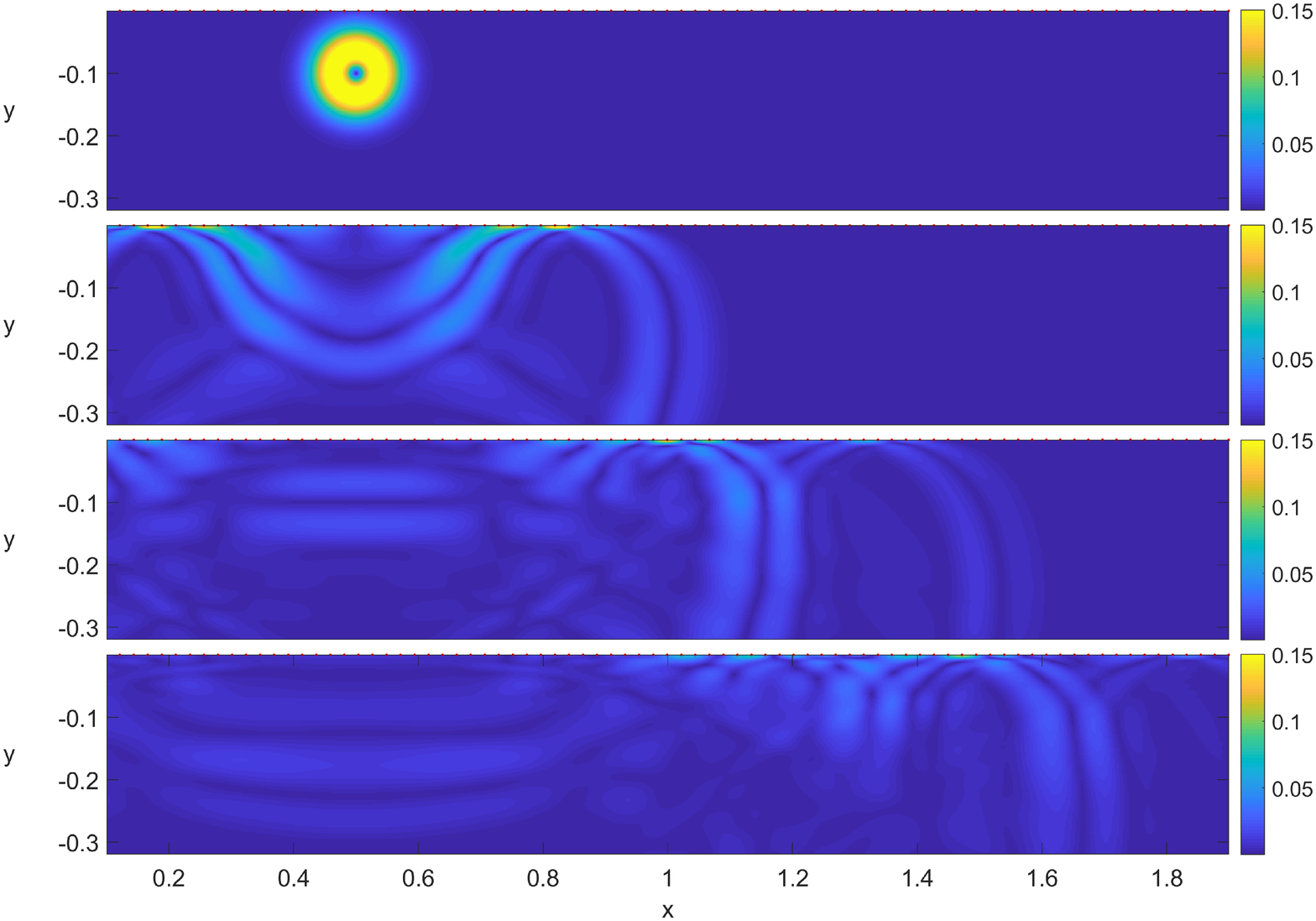}\\
\hspace*{-2cm}\includegraphics[width=1.2\textwidth]{fig6b.eps}

\caption{A simulation of the propagation of an initially circular wave through a medium having wavespeed increasing rapidly with depth is shown in the upper panel. The wavefront can be seen refracting back up to the surface. The top figure shows the initial disturbance at $t=0$. Each succeeding figure is $0.15$ seconds later than the previous. Distances are given in kilometres. A row of sensors is placed at the top of the space. In the lower panel, the velocity data for each of the sensors is shown.}
  %\end{center}
  \label{fig4}
\end{figure}

A final simulation is shown in Fig. \ref{fig4}, where the density and wavespeeds increase rapidly with depth.  These parameters are loosely based on the seismic structure of the upper level of an antarctic ice-sheet, in which the density in the packed snow near the top is quite low, but increases rapidly over a vertical distance of approximately 400 metres \cite{Reeh,Schlegel}. To approximate such a situation, we have set the nominal upper density to $\rho_1=300 kg/m^3$ and wavespeeds to $v_{p1}=0.4 km/s$ and $v_{s1}=0.3 km/s$. Lower nominal values are $\rho_2=900 kg/m^3$ with $v_{p2}=3.5 km/s$ and $v_{s2}=1.8 km/s$. These wavespeeds are related to the Lam\'e parameters by equation (\ref{wavespeeds}). These nominal values were used to create functions for density and wavespeeds increasing smoothly with depth. The vertical profile for density was set to $\rho(y)=\rho_2-(\rho_2-\rho_1)\exp(20y)$ with $-0.4<y<0$ in kilometres. The wavespeed profiles were set in the same way. Absorbing boundary layers were implemented on left and right, and on the bottom edge, but not on the top. This allows the top boundary to act as a reflecting surface, although in this simple example, we have not specifically implemented zero-normal-stress boundary conditions. The rapid change in wavespeed causes downward travelling signals generated near the surface to be refracted back up and thus to bounce along the underside of the surface at $y=0$. A row of sensors has been placed at the top of the space, and the $x$ and $y$ velocities recorded for each sensor at each integration step. The formulation of the numerical scheme calculates both displacements and velocities at every time step. In this case we show the velocities at the sensor points. This sensor data is shown in the lower panel, produced using the "wiggle" \textsc{Matlab} function \cite{wiggle}. This simulation was performed in a regularly spaced $x-y$ grid $1600 \times 320$ points.

In the top frame, the initial, circular disturbance is shown. This disturbance expands, but the variation in wavespeed causes the wavefront to distort, so that in the second frame, the front has already refracted back up to the top. The front then reflects off the surface at $y=0$, and proceeds in a downward direction, which again refracts back up to the top in the third frame and again in the fourth. This series of reflections is seen in the series of wavefronts detected by the sensors in the lower panel.

\section*{Technical Details}
All of the simulations were run on a computer employing an Intel I7-7700 CPU, 32 GB RAM and a Quadro GP100 GPU. The operating system was Ubuntu 18.04. The computations were performed using Fortran, compiled using the PGI Fortran compiler \cite{pgi}, with most of the calculations being carried out by the GPU. The displacements $u^X$ and $u^Y$ were written to file and later read into \textsc{Matlab} to produce the figures. All calculations were performed using double precision arithmetic. For the $560\times560$ grid calculations in this paper, the analytic calculations took only 7 seconds to run, this time being primarily for the calculations of the integrals (\ref{integrals}). The numerical calculations of Fig. \ref{fig1} to \ref{fig4} took 3-4 minutes. These consisted of 4 units of time, each unit using 280 iterations through the four steps of the Runge-Kutta algorithm. A much smaller number of steps may be used, and doing so produces results which are not noticeably different, but this number of steps was required to bring the difference between analytic and numerical models down to the level of the non-physical reflections (in our case, down to approximately 0.001). In all simulations the number of Fourier modes in each direction was one fifth of the number of grid points in that direction. The majority of required computations are for the solution of equation (\ref{matrixode}). By re-writing the sums in these equations as pure Matrix products, they may be performed in parallel using CUDA code for  multiplying matrices efficiently on the GPU. This code was adapted from the sample provided in the PGI Fortran compiler user guide \cite{pgi}, and significantly improved for the particular task described in this paper. These algorithmic changes reduced the runtime by at least an order of magnitude.
\section{Summary}

For the case of homogeneous media, an exact solution to the elastic wave equation in two dimensions has been developed, with general initial conditions. For initial conditions of some symmetry, along with infinite boundary conditions, an analytic function was found which describes the displacement components for any chosen time. This provides an excellent test for the accuracy of similar numerical methods.

By combining powerful numerical techniques, a new scheme has been implemented for forward integration of the elastic wave equation. The spectral (Fourier) method allows for derivatives to be calculated immediately and analytically from the basis functions. The classic fourth order Runge-Kutta method enables quick and accurate integration. Evaluation of massively parallel systems of equations is well suited to computation on modern graphics hardware, which is implemented using the chosen Fortran compiler.

Combining this numerical scheme with a suitable absorbing boundary layer allows for forward modelling of elastic waves. In the case of a homogeneous medium, the numerical method has been verified against the analytic solution. Waveforms from two inhomogeneous models are shown as illustrative examples.

This numerical method and its implementation on GPU architecture is presented for use by researchers where there is a need to investigate large numbers of relatively simple seismic models, tailored to environmental applications.  The ability to produce results rapidly using desktop computing and the agility in use allows for experimentation over a range of model parameters.

\dataccess{The Fortran and \textsc{Matlab} code for the simulations in this paper is available at \url{https://github.com/StephenJWalters/math-seismic}}

\aucontribute{SJW wrote the code, developed the numerical model and prepared the initial draft of the paper, LKF derived the analytic solution, AMR devised the overarching research program, and provided the seismological applications context. All authors were involved in the revision and approval of the final manuscript.}

\competing{We have no competing interests.}

\funding{This research was supported under Australian Research Council's Special Research Initiative for Antarctic Gateway Partnership (Project ID SR140300001), and Discovery Program (DP190100418).}

\ack{We are grateful to Professor Shaolin Liu and three anonymous referees, whose constructive comments have improved this paper considerably.}
%%%%%%%%%% Insert bibliography here %%%%%%%%%%%%%%

\section{Appendix}
After performing the integrals in equation (\ref{cf1}), the displacements are written for readability in the form
\begin{eqnarray}
u^X(x,y,t)&=&\sum_{m=\pm 1} \sum_{n=\pm 1} \sum_{p=\pm 1}\bigg( \frac{S_1+S_2+S_3+S_4+S_5}{8 D_1^4 D_2^2 D_3}+\frac{S_6 S_7}{4 D_1^5} \bigg ),\nonumber\\
&&\text{where the auxiliary functions are} \nonumber\\
&&S_1=m\bigg[3 i n\beta t \left(2-\beta ^2 t^2\right)+2 \beta^4 t^4-3 \beta^2 t^2+r^4-8 r^2+4 \bigg]\nonumber \\
&&S_2=n \beta^3 t^3 (4 x-m p y)\nonumber\\
&&S_3=3 \beta ^2 t^2 \left(m x^2-m y^2+p x y+i m p y-3 i x\right)\nonumber\\
&&S_4=n\beta t [2 x^3-x y^2-13 x+m p y(4 x^2+y^2 -5)-2 i (3 p x y+5 m x^2+m y^2)]\nonumber\\
&&S_5=-8 p x y-4 i (r^2-2)(x+m p y)\nonumber\\
&&S_6=3 n \beta t (x-m i)\nonumber\\
&&S_7=\text{ArcTanh}(\frac{i-p y}{D_1})+\text{ArcTanh}(\frac{n \beta t -m x+p y}{D_1})\nonumber\\
&&D_1=\sqrt{r^2-\beta ^2 t^2-2-2i(m x+p y)}\nonumber\\
&&D_2=n \beta t+m x-i\nonumber\\
&&D_3=n \beta t+p y-i\nonumber\\
&&r=\sqrt{x^2+y^2}\nonumber
\end{eqnarray}
Unsurprisingly, the solution for $u^Y$ is identical to that for $u^X$ but with $(x,y)$ replaced by $(y,-x)$:
\begin{eqnarray}
u^Y(x,y,t)&=&\sum_{m=\pm 1} \sum_{n=\pm 1} \sum_{p=\pm 1} \bigg (\frac{T_1+T_2+T_3+T_4+T_5}{8 E_1^4 E_2^2 E_3}+\frac{T_6 T_7}{4 E_1^5}\bigg ),\nonumber\\
&&\text{where} \nonumber\\
&&T_1=m\bigg[3 i n\beta t \left(2-\beta ^2 t^2\right)+2 \beta^4 t^4-3 \beta^2 t^2+r^4-8 r^2+4 \bigg]\nonumber \\
&&T_2=n \beta^3 t^3 (4 y+m p x)\nonumber\\
&&T_3=3 \beta ^2 t^2 \left(m y^2-m x^2-p x y-i m p x-3 i y\right)\nonumber\\
&&T_4=n\beta t [2 y^3-y x^2-13 y-m p x(4 y^2+x^2-5)-2 i(5 m y^2+m x^2-3 p x y)]\nonumber\\
&&T_5=8 p x y-4 i (r^2-2)(y-m p x)\nonumber\\
&&T_6=3 n \beta t (y-m i)\nonumber\\
&&T_7=\text{ArcTanh}(\frac{i+p x}{E_1})+\text{ArcTanh}(\frac{n \beta t -m y-p x}{E_1})\nonumber\\
&&E_1=\sqrt{r^2-\beta ^2 t^2-2-2i(m y-p x)}\nonumber\\
&&E_2=n \beta t+m y-i\nonumber\\
&&E_3=n \beta t-p x-i\nonumber\\
&&r=\sqrt{x^2+y^2}\nonumber
\end{eqnarray}


\vskip2pc

\begin{thebibliography}{9}
\bibitem{aki} Aki K, \& Richards PG. 2002 \textit{Quantitative seismology.}


\bibitem{kennett} Kennett BL. 2001 \textit{The Seismic Wavefield: Volume 1, Introduction and Theoretical Development.} Cambridge University Press.

\bibitem{lamb}  Lamb H. 1904 \textit{On the propagation of tremors over the surface of an elastic solid.} Phil. Trans. R. Soc. Lond. A 203, 1-42.

\bibitem{garvin} Garvin WW. 1956 \textit{Exact transient solution of the buried line source problem.} Proc. R. Soc. A 234(1199), 528-541.

\bibitem{kausel} Kausel E. 2013 \textit{Lamb's problem at its simplest}. Proc. R. Soc. A, 469(2149), doi: 10.1098/rspa.2012.0462.

\bibitem{gosselin} Gosselin-Cliche B, \& Giroux B. 2014 \textit{3D frequency-domain finite-difference viscoelastic-wave modeling using weighted average 27-point operators with optimal coefficients.} Geophysics 79(3), T169-T188.

\bibitem{carcione} Carcione JM. 1993 \textit{Seismic modeling in viscoelastic media.} Geophysics 58(1) 110-120.

\bibitem{diaz} Diaz J, \& Ezziani A. 2010 \textit{Analytical solution for waves propagation in heterogeneous acoustic/porous media. Part I: the 2D case.} Communications in Computational Physics, 7(1), 171.

\bibitem{igel} Igel H. 2017 \textit{Computational seismology: a practical introduction.} Oxford University Press.

\bibitem{fichtner} Fichtner A, Igel H, Bunge HP, \& Kennett BL. 2009 \textit{Simulation and inversion of seismic wave propagation on continental scales based on a spectral-element method.} JNAIAM 4(1-2), 11-22.

\bibitem{koma} Komatitsch D \& Vilotte JP. 1998 \textit{The spectral element method: an efficient tool to simulate the seismic response of 2D and 3D geological structures.} Bulletin of the seismological society of America, 88(2), 368-392.

\bibitem{maeda} Maeda T, Takemura S \& Furumura T. 2017 \textit{OpenSWPC: an open-source integrated parallel simulation code for modeling seismic wave propagation in 3D heterogeneous viscoelastic media.} Earth, Planets and Space, 69(1), 102.

\bibitem{sens} Sens-Sch{\"o}nfelder C, \& Wegler U. 2011 \textit{Passive image interferometry for monitoring crustal changes with ambient seismic noise.} Comptes Rendus Geoscience, 343(8-9), 639-651.

\bibitem{tsai} Tsai VC, Minchew B, Lamb MP \& Ampuero JP. 2012 \textit{A physical model for seismic noise generation from sediment transport in rivers.} Geophysical Research Letters, 39(2).

\bibitem{winberry} Paul Winberry J, Anandakrishnan S, Wiens DA, \& Alley RB. 2013 \textit{Nucleation and seismic tremor associated with the glacial earthquakes of Whillans Ice Stream, Antarctica.} Geophysical Research Letters, 40(2), 312-315.

\bibitem{walters} Walters SJ \& Forbes LK. 2019 \textit{Fully 3d Rayleigh-Taylor instability in a Boussinesq fluid.} ANZIAM J. 61(3), 286-304.

\bibitem{boyd}Boyd JP. 2001 \textit{Chebyshev and Fourier spectral methods.} Dover, New York.


\bibitem{nm} Nelder JA \& Mead R. 1965 \textit{A simplex method for function minimization}. The Computer Journal, 7(4), 308-313.

\bibitem{assi} Assi H, \& Cobbold RS. 2017 \textit{Compact second-order time-domain perfectly matched layer formulation for elastic wave propagation in two dimensions.} Mathematics and Mechanics of Solids, 22(1), 20-37.

\bibitem{assi2} Assi H. 2016 \textit{Time-domain modeling of elastic and acoustic wave propagation in unbounded media, with application to metamaterials} (Doctoral dissertation, University of Toronto, Canada).

\bibitem{atkinson} Atkinson KE. 1978 \textit{An introduction to numerical analysis} John Wiley \& Sons.

\bibitem{Reeh}Reeh N, Fisher DA, Koerner RM, \& Clausen HB. 2005  \textit{An empirical firn-densification model comprising ice lenses}. Annals of Glaciology, 42, 101-106.

\bibitem{Schlegel}Schlegel R, Diez A, L{\"o}we H, Mayer C, Lambrecht A, Freitag J, Miller H, Hofstede C \& Eisen O. 2019 \textit{Comparison of elastic moduli from seismic diving-wave and ice-core microstructure analysis in Antarctic polar firn}. Annals of Glaciology, 60(79), 220-230.%doi: 10.1017/aog.2019.10


\bibitem{wiggle} Portugal R. `https://au.mathworks.com/matlabcentral/fileexchange/38691-wiggle'. Accessed 18 September 2019.

\bibitem{pgi} PGI Community Edition. `https://www.pgroup.com/products/community.htm'. Accessed 14 May 2019.

\end{thebibliography}
\end{document}